\documentclass[aps,prc,twocolumn,amsmath,amssymb,nofootinbib]{revtex4-1}
\usepackage{graphicx,epsfig}
\usepackage{amsmath,braket,bm}
\usepackage {color}
\usepackage[bookmarks=true]{hyperref}
\hypersetup{
	colorlinks=true,
	pdfborder={0 0 0},
	citecolor=blue,
	linkcolor=blue,
	urlcolor=blue
}

\begin{document}
\title{Imprint of nuclear bubble in nucleon-nucleus diffraction}
\author{V. Choudhary$^{1}$}
\email{vchoudhary@ph.iitr.ac.in}
\author{W. Horiuchi$^{2}$}
\email{whoriuchi@nucl.sci.hokudai.ac.jp}
\author{M. Kimura$^{2,3}$}
\email{masaaki@nucl.sci.hokudai.ac.jp}
\author{R. Chatterjee$^{1}$}
\email{rchatterjee@ph.iitr.ac.in}

\affiliation{$^{1}$Department of Physics, Indian Institute of Technology Roorkee, Roorkee 247 667, India}

\affiliation{$^2$Department of Physics, Hokkaido University, 060-0810 Sapporo, Japan}

\affiliation{$^3$Nuclear Reaction Data Centre, Faculty of Science, Hokkaido University, 060-0810 Sapporo, Japan}

	
\begin{abstract}
\begin{description}
\item[Background]
  The density of most nuclei is constant in the central region
  and is smoothly decreasing at the surface. 
  A depletion in the central part of the nuclear density can have
  nuclear structure effects leading to the formation of ``bubble'' nuclei.
  However, probing the density profile of the nuclear interior is,
  in general, very challenging.
\item[Purpose]	
  The aim of this paper is to investigate the nuclear bubble structure,
  with nucleon-nucleus scattering, and quantify the effect that has
  on the nuclear surface profile. 
\item[Method]
  We employed high-energy nucleon-nucleus scattering
  under the aegis of the Glauber model to analyze  
  various reaction observables, which helps in quantifying
  the nuclear bubble. The effectiveness of this method 
  is tested on $^{28}$Si with harmonic-oscillator (HO) densities,
  before applying it on even-even $N=14$ isotones,
  in the $22\leq A \leq 34$ mass range, with realistic densities obtained from
  antisymmetrized molecular dynamics (AMD).	
 \item[Results]
   Elastic scattering differential cross sections
   and reaction probability for the  proton-$^{28}$Si reaction
   are calculated using the HO density to design tests for signatures
   of nuclear bubble structure. We then quantify the degree of bubble structure 
   for $N=14$ isotones with the AMD densities by analyzing
   their elastic scattering at 325, 550 and 800 MeV incident energies.
   The present analyses suggest $^{22}$O as a  candidate for a bubble 
   nucleus, among even-even $N=14$ isotones,
   in the $22\leq A \leq 34$ mass range.
\item[Conclusion]
  We have shown that the bubble structure information
  is imprinted on the nucleon-nucleus elastic scattering 
  differential cross section, especially in the first diffraction peak.
  Bubble nuclei tend to have a sharper nuclear surface
  and deformation seems to be a hindrance in their emergence. 	
\end{description}
    
 \end{abstract}
	
\maketitle
\section{introduction}
Advances of radioactive beam facilities have allowed us
to study nuclei with extreme neutron to proton ratios. In fact, 
close to the neutron drip line, one has discovered exotic features
like haloes~\cite{Tanihata85,Tanihata13}
- an extended low density tail in the neutron matter distribution. 
At least for light nuclei, this was thought to be a threshold phenomenon 
resulting from the presence of a loosely bound state near the continuum. 
In this context, with current interest moving towards the medium mass region, 
another exotic structure that of a depression in the central part 
of nuclear density - called a ``bubble'' - has attracted considerable attention.

Systematic studies of electron scattering
of stable nuclei have revealed
that the central density of stable nuclei is almost constant,
$\rho_{0} \approx 0.16$  fm$^{-3}$~\cite{Hofstadter}.
In light nuclei, distinct nuclear orbitals
play a role in the emergence of the bubble structure.
If the $s$-orbitals are empty,  the interior density
of nuclei becomes depleted. For example, Refs.~\cite{Grasso,Li} showed
that the central depression of the proton density in $^{34}\textrm{Si}$
is about 40\% as compared to stable $^{36}\textrm{S}$ using several
mean-field approaches, originated from the proton deficiency
in the $1s_{1/2}$ orbit. The possibility of forming  bubble nuclei
have also been explored theoretically
in the medium~\cite{Campi,Davis,Khan,Grasso},
and superheavy mass regions~\cite{Bender}. 

The experimental indication of the central depression of protons
in the unstable nucleus $^{34}\textrm{Si}$
was recently reported using $\gamma$-ray spectroscopy~\cite{mutschler}.
Electron scattering on unstable nuclei
is the most direct way to probe the central depression of
proton density in bubble nuclei.
Recently, the SCRIT electron scattering facility
has succeeded in extracting information
about the nuclear shape of $^{132}\textrm{Xe}$~\cite{Tsukada}.

However, unlike a hadronic probe,
which is sensitive to both neutrons and protons,
the electron scattering has difficulty
to probe the neutron density distribution even for stable nuclei~\cite{PREX}.
In this context, it is worth mentioning that proton-nucleus scattering has been successfully applied  
to deduce the nuclear matter density distributions~\cite{Sakaguchi}.
Proton scattering can also be extended for unstable nuclei
with the use of inverse kinematics measurement as demonstrated
in Ref.~\cite{Matsuda13}. Indeed, this motivates us to inquire
if information on the bubble structure in nuclei
can be investigated with nucleon-nucleus scattering.

In this paper, we perform a systematic study to test the nucleon-nucleus scattering as a probe 
for the  nuclear bubble structure. This paper is organized as
follows. Section~\ref{reaction.sec} briefly presents the formalism that describes the
nucleon-nucleus collision at high incident energy within the Glauber model, wherein the
elastic scattering and total reaction cross sections are evaluated.  Using this
formulation, in Sec.~\ref{bubble.sec},  we discuss how signatures of the nuclear bubble
structure are reflected in the cross sections by using an example of a simple ideal
case, $^{28}$Si. We show the relationship between the internal depression and the
surface diffuseness, and propose a practical way to evaluate the bubble structure. 
For this purpose, the generalized ``bubble'' parameter is introduced as a measure of the nuclear bubble structure. 
In this work, we also examine the structure of $N=14$ isotones,
$^{22}$O, $^{24}$Ne, $^{26}$Mg, $^{28}$Si, $^{30}$S, $^{32}$Ar, and $^{34}$Ca.
Section~\ref{isotones.sec} presents
details of the structure calculation by the antisymmetrized molecular
dynamics (AMD) model. The formalism is briefly explained in Sec.~\ref{amd.sec},
and the resulting structure information focusing on
the bubble structure is given in the following Sec.~\ref{structure.sec}.
Section~\ref{discussion.sec} demonstrates how the nucleon-nucleus
scattering works for extracting the bubble parameter
of the nuclear density distributions. We discuss the feasibility
through a systematic analysis of the 
elastic scattering differential cross sections
with various density profiles.
The conclusions of our study are presented in Sec.~\ref{conclusions.sec}. 
Some details on how nuclear structure parameters are evaluated in the AMD 
are in appendix A.

\section{Nucleon-nucleus reactions with Glauber model}
\label{reaction.sec}

The Glauber theory offers a powerful description
of high-energy nuclear reactions~\cite{glauber}.
Here we consider the normal kinematics in
which the incident proton is bombarded on a target nucleus.
Thanks to the eikonal and adiabatic approximations,
the final state wave function of the target nucleus
after the collision is simplified as	
\begin{equation}
\left|\phi_{f}\right> = e^{i\chi}\left|\phi_{i}\right>,
\end{equation}	
where $\left|\phi_{i}\right>$ represents the initial wave function
of the target nucleus,
and $e^{i\chi}$ is the phase-shift function, which
includes all the information about the nucleon-nucleus collision.
The elastic scattering amplitude for the nucleon-nucleus
reaction is given by	
\begin{equation}
  F(\bm{q}) = \dfrac{iK}{2 \pi}\int d\bm{b}\,e^{i\bm{q}\cdot\bm{b}}
  (1-e^{i\chi_{N}(\bm{b})}),
\end{equation}
where $K$ is the relative wave number of the incident nucleon, $\bm{b}$
is the impact parameter vector perpendicular to the beam direction,
and $\bm{q}$ is the momentum transfer vector of the incident nucleon.
With this scattering amplitude,
the elastic scattering differential cross section can be evaluated by
\begin{equation}
\frac{d\sigma}{d\Omega} = |F(\bm{q})|^{2}.
\end{equation}
The total reaction cross section of the nucleon-nucleus collision
can be calculated by
\begin{equation}
\sigma_{R} = \int d\bm{b}\,P(\bm{b})
\label{rcs.eq}
\end{equation}   
with the nucleon-nucleus reaction probability defined as
\begin{equation}    
 P(\bm{b})  =  1-|e^{i\chi(\bm{b})}|^{2}.
\label{reacprob.eq}
\end{equation}

Since the evaluation of the phase-shift function is  demanding in general,
for the sake of simplicity we employ the optical-limit approximation (OLA).
As presented in Refs. \cite{Varga02,Nagahisa18,Ibrahim09,Hatakeyama18},
the OLA works well for many cases of the proton-nucleus scattering so
that the multiple scattering effects can be ignored.
The optical phase-shift function for the nucleon-nucleus
scattering in the OLA is given by 	
\begin{equation}
  e^{i\chi_N(\bm{b})} \approx \exp\left[
  -\int d\bm{r} \rho_{N}(\bm{r}) \Gamma_{NN}(\bm{b}-\bm{s})\right],
\end{equation}
where $\bm{r} = (\bm{s},z)$,  and $\bm{s}$ is
the two-dimensional vector perpendicular to the beam direction $z$.
$\rho_{N}(\bm{r})$ denotes the nucleon density distributions
measured from the center of mass of the system. The crux of any calculation will be 
to calculate this density with reliable nuclear structure models. This is also the primary point  
where information on the bubble structure enters into the Glauber model and is reflected in 
the scattering or reaction observables.

$\Gamma_{NN}$ is the profile function, which describes the
nucleon-nucleon collisions.
The profile function for the nucleon-nucleon scattering
is usually parametrized as~given in Ref.~\cite{Lray}	
\begin{equation}
  \Gamma_{NN}(\bm{b}) = \dfrac{1-i\alpha_{NN}}{4 \pi \beta_{NN}}
  \sigma_{NN}^{\rm tot}\exp\bigg(-\dfrac{\bm{b}^2}{2 \beta_{NN}}\bigg),
\end{equation}	
where $\alpha_{NN}$ is the ratio of the real part
to the imaginary part of the nucleon-nucleon scattering amplitude
in the forward direction,
$\beta_{NN}$ is the slope parameter of the differential cross section,
and $\sigma_{NN}^{\rm tot}$ is the nucleon-nucleon total cross section.
Standard parameter sets of the profile function
are listed in Refs.~\cite{Horiuchi07,Ibrahim08}.

\section{How is nuclear bubble structure reflected?}
\label{bubble.sec}
In this section, we discuss how the nuclear bubble gets reflected in the proton-nucleus
scattering at high incident energies, where the Glauber model works fairly well. For the
sake of simplicity, we use the averaged $NN$ profile function given in
Ref.~\cite{Horiuchi07} and ignore the Coulomb interaction.
Note that the difference between the $pp$ and $pn$ cross sections in
the profile functions can be neglected
in the total reaction cross section calculations
at the incident energy of $E \gtrsim 300$ MeV~\cite{Nagahisa18}.

\subsection{Density distribution of $^{28}$Si}
\label{density}
Here, we discuss the density distribution of  $^{28}$Si
within the harmonic-oscillator (HO) model. 
First, we consider two types of configurations, $(0d)^{12}$ and $(0d)^8(1s)^4$, and
calculate their density distributions with the center-of-mass
correction~\cite{Ibrahim09}, which are denoted as $\rho^d( r)$ and $\rho^s( r)$,
respectively. Note that $\rho^d(r)$ shows
the most prominent bubble structure because of the
vacancy of the $1s$-orbit, while $\rho^s( r)$ does not.
Then, we interpolate these two densities as
\begin{equation}
  \rho(\alpha;r)= (1-\alpha)\rho^{d}(r)  +\alpha \rho^{s}(r),
\label{HOdens.eq}
\end{equation}
where the mixing parameter $\alpha$ ($0\leq\alpha\leq 1$) controls the occupation
probability of the $1s$-orbit. Consequently, $\alpha=0$ yields
the most bubbly density,
whereas $\alpha=1$ yields non-bubble density. For a given value of $\alpha$, the size parameter of
HO is chosen to reproduce the observed point-proton root-mean-square (rms) radius, 3.01
fm~\cite{Angeli13}. 

To quantify a degree of ``bubble'', we introduce the bubble parameter ($G$) as, 
\begin{equation}
G = \dfrac{\rho(D)-\rho(0)}{\rho(D)},
\label{bubblepara.eq}
\end{equation}
where, $D$ denotes the reference radius at which the $\rho^{d}(r)$ or $\rho(\alpha=0;
r)$ takes its maximum value. In the case of $^{28}{\rm Si}$, $D=1.8$ fm.
$\rho(0)$ and $\rho(D)$ represent the densities at $r=0$ and $D$, respectively.  
We remark that this is an extension of the bubble parameter (depletion fraction)
given in Ref.~\cite{Grasso}, where it is defined only by positive values. This extension
enables us to quantify the degree of the bubble structure for any nuclear density
distribution irrespective of whether it exhibits a bubble or not.

\begin{figure}[t]
\begin{center}
\epsfig{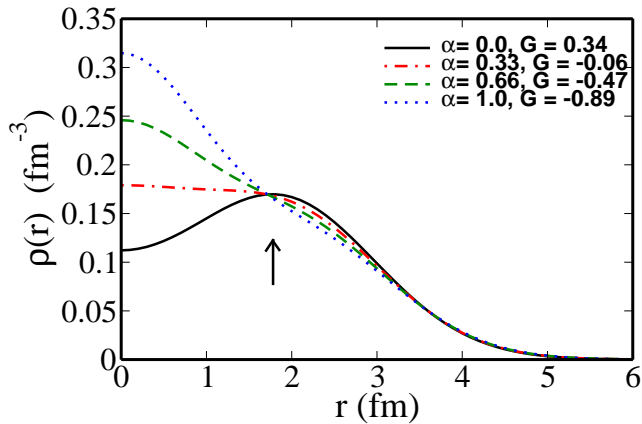}
\caption{Matter density distributions of $^{28}$Si with various bubble parameters
 ($G$).   The arrow indicates the reference radius, 1.8 fm.   See text for more
  details.}\label{dens28Si.fig}                 
\end{center}
\end{figure}

Figure~\ref{dens28Si.fig} displays how the matter density distribution of $^{28}$Si and 
the corresponding $G$ value change depending on the mixing parameter $\alpha$. 
In the present case of $^{28}$Si, the values of $G$ range from 0.34 ($\alpha=0$) to
$-$0.89 ($\alpha=1$), allowing for negative values which  signify that the central density
is higher than the density at the reference radius.
Apparently, the bubble degree is maximized at $G=0.34$ with $\alpha=0$, which clearly
exhibits a strong depression of the central density, thereby suggesting the bubble
structure. The value of $G$ decreases with increasing the mixing of the $1s$-orbits. 
An almost flat behavior of the density distribution ($G\approx 0$) is obtained with $\alpha=0.33$. 

In the same manner, we construct the model densities for other $N=14$ isotones from
$^{22}{\rm O}$ to $^{34}{\rm Ca}$. We calculate $\rho^{d}(r)$ and $\rho^{s}(r)$ as the
density distributions of the $(0d)^{A-16}$ and $(1s)^4(0d)^{A-20}$ configurations with
$22\leq A \leq 34$. These two densities are interpolated as in Eq. (\ref{HOdens.eq})
and used for the reaction calculation in the following sections. The reference radius
$D$ and the bubble parameters are also defined in the same way. 

\subsection{Bubble structure in proton-$^{28}$Si reactions}

\begin{figure}[ht]    
\begin{center}
\epsfig{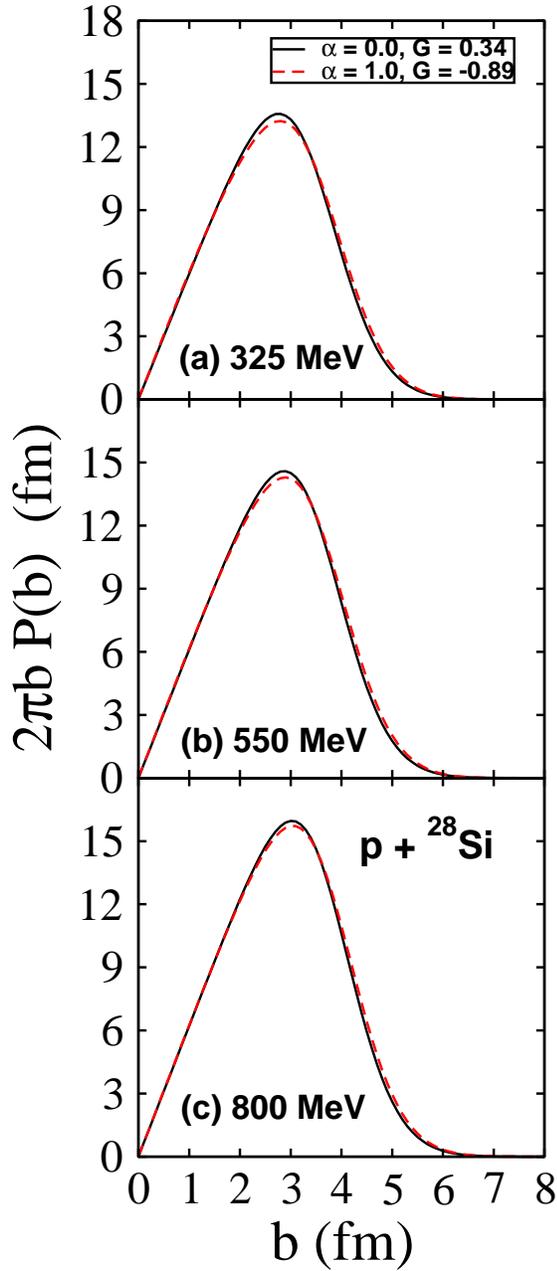}
\caption{Reaction probabilities multiplied by $2\pi b$ for proton-$^{28}\textrm{Si}$ reactions
  at (a) 325, (b) 550, and (c) 800 MeV.   The configurations with $\alpha=0$ ($G=0.34$) and $1$
  ($G=-0.89$)   are employed.}\label{reacprob.fig}
\end{center}
\end{figure} 

How are the different density profiles displayed in Fig.~\ref{dens28Si.fig} reflected in
the reaction observables? To address this question, we calculated the reaction
probability $P(b)$ given in Eq.~(\ref{reacprob.eq}), which is the integrand of the total
reaction cross section [Eq.~(\ref{rcs.eq})]. Figure~\ref{reacprob.fig} shows the
reaction probability multiplied by $2\pi b$ for proton-$^{28}\textrm{Si}$ scattering as
a function of the impact parameter. The density distributions with $\alpha=0$
($d$-dominance, $G=0.34)$ and 1 (maximum $s$ configuration, $G=-0.89)$ are examined to
see the difference between the two extreme configurations. 
The reaction probabilities for these configurations
are almost identical at small impact parameters up
to $\approx 2$ fm despite the fact that these two density profiles
show significant difference in Fig.~\ref{dens28Si.fig}.
Since the nucleon-nucleon interaction is large
enough, the reaction occurs almost entirely in the internal region 
below the nuclear radius as predicted by the black sphere model, 
which explains high-energy proton-nucleus scattering 
fairly well~\cite{Kohama04,Kohama05,Kohama16}. 
Therefore, it cannot directly probe the internal part
of the nuclear density profile.
 
However, we see some differences beyond $b\approx 2$ fm at the nuclear surface, which
suggests the possibility for extracting the surface information from the cross
sections. In fact, the relation between elastic scattering differential cross sections
and nuclear surface diffuseness has been discussed in Ref.~\cite{Amado} and recently in  
Ref.~\cite{Hatakeyama18}. It was shown that the smaller the nuclear diffuseness, the
larger is the cross section at the peak position of the first diffraction. Note that 
the bubble density distribution
has smaller diffuseness than the non-bubble density distribution as seen in
Fig.~\ref{dens28Si.fig}, because the former ($\alpha=0$) includes only the
$d$-wave configuration, while the latter ($\alpha=1$) includes the
$1s_{1/2}$ configuration which has a longer tail in the asymptotic region.
Consequently, we expect that the bubble structure
gives the larger cross section at the first diffraction peak.

\begin{figure}[ht]    
\begin{center}
\epsfig{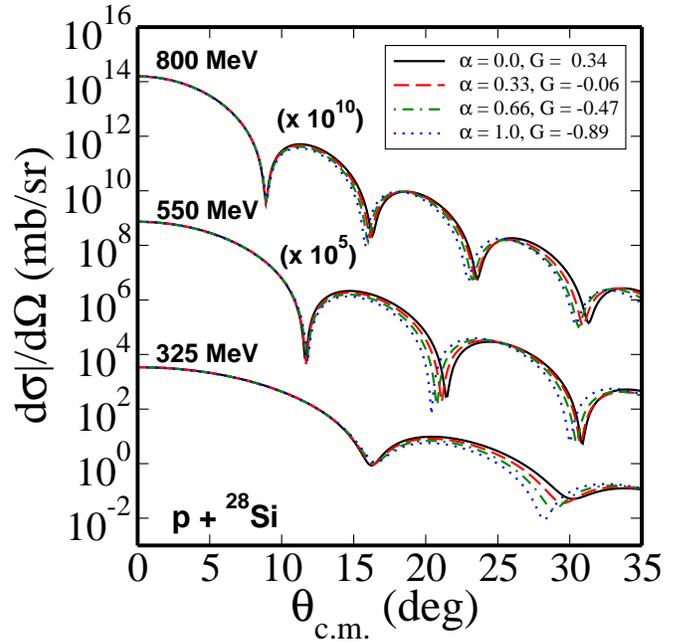}
\caption{Elastic scattering differential cross sections
  of proton-$^{28}\textrm{Si}$ reactions
  at 325, 550, and 800 MeV with various bubble parameters.}\label{dcs28Si.fig}
\end{center}
\end{figure}

To confirm this numerically, we calculated the proton-$^{28}$Si elastic scattering
differential cross sections. Figure~\ref{dcs28Si.fig} plots the elastic scattering
differential cross sections of proton-$^{28}\textrm{Si}$ reactions  at 325, 550, and 800
MeV with various bubble parameters. As expected, the cross section at the first peak
position is largest for the ideal bubble configuration with $\alpha=0$ and it decreases
with increasing $\alpha$ and decreasing $G$. This suggests a practical way to identify
the bubble structure using a hadronic probe.

\section{Bubble structure of $N=14$ isotones}
\label{isotones.sec}
We have seen that the difference between the bubble and non-bubble nuclei can be
detected in the elastic scattering differential cross sections. To demonstrate the
feasibility of this idea, we take the density distributions obtained from a microscopic
structure model, the AMD,  and try to extract the information on the nuclear bubble from
the reaction observable. Here the ground-state density distributions of $N=14$ isotones
are examined as they exhibit the bubble structure in its isotone chain~\cite{Grasso}.  

\subsection{Framework of AMD} 
\label{amd.sec}

The AMD~\cite{AMDRev1,AMDRev2} is a fully microscopic approach
and offers a non-empirical description of light to medium nuclei.
Here we briefly explain how we obtain the density distributions
for the $N=14$ isotones within the AMD framework.  
The Hamiltonian for a nucleon system with the mass number $A$ is given by
\begin{align}
 H = \sum_{i}t(i) - T_{\rm cm} + \sum_{ij} v_{NN}(ij),
\end{align}
where $t(i)$ is the kinetic energy of the single nucleon
and the center-of-mass kinetic energy $T_{\rm cm}$
is exactly removed. The Gogny D1S parameter set~\cite{GognyD1S}
is employed as a nucleon-nucleon effective interaction $v_{NN}$,
which is known to give a fairly good description
for this mass region \cite{Sumi12, Wata14, Peru14}.

The variational basis function of the AMD is represented
by a Slater determinant projected to the positive-parity state as 
\begin{align}
 \Phi = \frac{1+P_x}{2}\mathcal{A}\left\{\varphi_1,...,\varphi_A\right\},
\end{align}
where $P_x$ is the parity operator, and
$\varphi_i$ is a Gaussian nucleon wave packet defined by 
\begin{align}
  \varphi_i=&\prod_{\sigma=x,y,z}\left(\frac{2\nu_\sigma}{\pi}\right)^{1/2}
 \exp\set{-\nu_\sigma\left(r_\sigma - Z_{i\sigma}\right)^2}\nonumber\\
 &\times(\alpha_i\chi_\uparrow + \beta_i \chi_\downarrow)
 (\ket{p} {\text or} \ket{n}).
\end{align}
The centroids $\bm Z$ and width $\bm \nu$ vectors of the Gaussian
and the spin variables $\alpha_i$ and $\beta_i$ are
the variational parameters. They are determined by the the frictional
cooling method~\cite{cooling} in such a way to minimize
the energy of the system under the constraint on the quadrupole
deformation parameter $\beta$. 

To describe the ground state of the $N=14$ isotones, the wave functions obtained by the frictional
cooling method are projected to the angular momentum $J=0$ and superposed employing  $\beta$ as a
generator coordinate (generator coordinate method; GCM~\cite{Hill53}),
\begin{align}
 \Psi_{0} = \sum_{i}g_iP^{J=0}\Phi(\beta_i), \label{eq:gcmwf}
\end{align}
where $P^{J=0}$ represents the angular momentum projector and the amplitudes $g_i$ are
determined by the diagonalization of the Hamiltonian. 
In the present study, the value of $\beta$ is chosen from 0.0 to 0.6 with an interval of 0.025.
The deformation parameter $\gamma$ is determined variationally,
and hence it takes an optimal value for
each $\Phi(\beta_i)$. Finally, the ground-state density distribution
is calculated as
\begin{align}
  \rho(\bm r) = 
  \frac{\braket{\Psi_{0}|\sum_{i=1}^A \delta^3(\bm r_i - \bm r_{\rm cm}-\bm r)|\Psi_{0}}}
  {\braket{\Psi_{0}|\Psi_{0}}}.
\end{align}
Note that the resulting density distribution is free from
the center-of-mass coordinate $\bm r_{\rm cm}$.
We also evaluate the quadrupole deformation parameters and
occupation probabilities of the $1s$-orbit according
to the procedure described in the appendix \ref{app:amd}.

\subsection{Density distributions of $N = 14$ isotones}
\label{structure.sec}
\begin{figure*}[th]
 \begin{center}    
  \includegraphics[width=0.9\hsize]{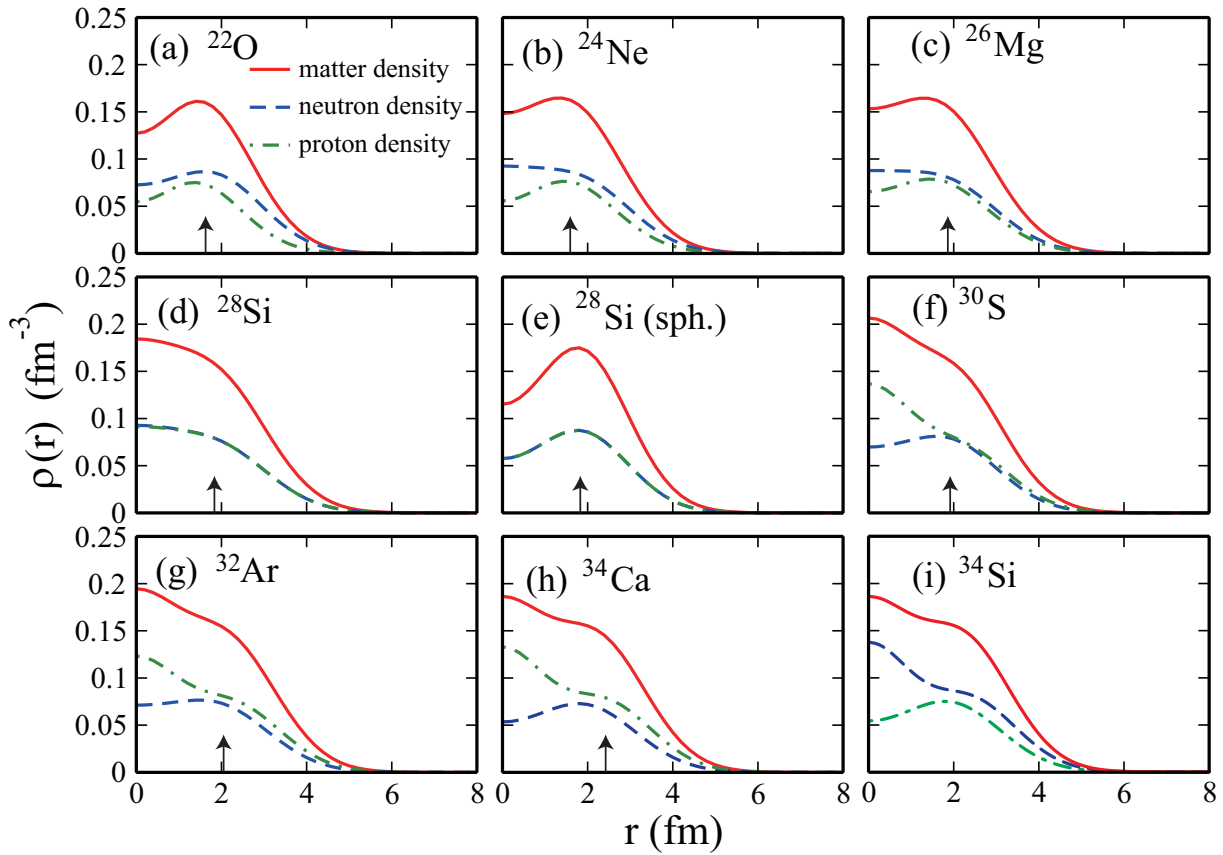}
  \caption{Density distributions of $N = 14$ isotones,
  (a) $^{22}$O, (b) $^{24}$Ne, (c) $^{26}$Mg, (d) $^{28}$Si,
  (e) $^{28}$Si (spherical), (f) $^{30}$S, (g) $^{32}$Ar, and
  (h) $^{34}$Ca obtained from the AMD wave function.
  The arrows indicate the reference radii for each isotone.
  See text for details. The density distribution of $^{34}{\rm Si}$ which is the mirror
  nucleus of $^{34}{\rm Ca}$ and has prominent
  proton bubble structure is also shown in the panel (i) for comparison.}
  \label{density.fig}
 \end{center}
\end{figure*}

Figure~\ref{density.fig} plots the matter, neutron and proton density distributions of
$N=14$ isotones obtained with the AMD. The rms matter radii,
quadrupole deformation parameters, 
and occupation probabilities of the $1s$-orbit are summarized in
Table~\ref{tab:deformation}.
The bubble parameters $G_{\rm AMD}$ were also calculated
from the AMD densities using Eq.~(\ref{bubblepara.eq}),
where the reference radius ($D$),
for each isotone, was derived from an effective density $\rho^d(r)$.
This $\rho^d(r)$ is essentially a HO density [as in Sec.~(\ref{density})] whose size parameter is adjusted so as to reproduce the rms matter radius, for each isotone, as obtained from the AMD density.

\begin{table}[bht]\caption{Rms matter radii $r_{m}$
    and the quadrupole deformation parameters $\beta,\gamma$,
    and neutron (proton) occupation probabilities of the $1s$-orbit
    $P_{1s}(n) [P_{1s}(p)]$ for $N = 14$ isotones obtained by the AMD.
    The bubble parameter $G_{\rm AMD}$ are extracted from the matter
    density distributions
 shown in Fig.~\ref{density.fig}.
 See text for more details.} 
 \label{tab:deformation}   
 \begin{center}
  \begin{ruledtabular}
   \begin{tabular}{lcccccr}
  &  $r_{m}$ (fm) & $\beta$ & $\gamma$ & $P_{1s}(n)$ & $P_{1s}(p)$ & $G_{\rm AMD}$\\
     \hline
    $^{22}$O  &     2.90 & 0.20 & $60^\circ$ & 0.21 & 0.01 & 0.21 \\
    $^{24}$Ne &     2.97 & 0.37 & $60^\circ$ & 0.28 & 0.01 & 0.09  \\
    $^{26}$Mg &     3.06 & 0.40 & $37^\circ$ & 0.26 & 0.05 & 0.04  \\
    $^{28}$Si &     3.11 & 0.40 & $60^\circ$ & 0.29 & 0.29 & $-$0.16 \\
    $^{28}$Si(sph.)&2.98 & 0.00 & --        & 0.01 & 0.01 & 0.34  \\
    $^{30}$S  &     3.11 & 0.27 & $43^\circ$ & 0.17 & 0.65 & $-$0.29 \\
    $^{32}$Ar &     3.21 & 0.27 & $60^\circ$ & 0.21 & 0.62 & $-$0.25 \\
    $^{34}$Ca &     3.26 & 0.12 & $60^\circ$ & 0.06 & 0.91 & $-$0.05  \\
   \end{tabular}
  \end{ruledtabular}
 \end{center}
\end{table}

We clearly see a prominent bubble structure in $^{22}$O in
which both the proton and neutron density distributions
exhibit depressed central
densities. Consequently, it has the largest bubble parameter among the $N=14$ isotones.
This is due to the almost spherical closed-shell configuration of this nucleus and the
resultant small occupation probabilities of the $1s$-orbit.
As the proton number increases, the nuclear quadrupole deformation becomes strong,
which mixes the $s$-, $d$- and $g$-orbits and effectively increases the occupation
probabilities of the $1s$-orbit. As a result, the bubble structure in the matter
density distributions is weakened in $^{24}{\rm Ne}$ and $^{26}{\rm Mg}$, and diminished
in $^{28}{\rm Si}$ which is most strongly deformed among the $N=14$
isotones. Indeed, Table~\ref{tab:deformation} shows that the bubble parameter
strongly correlates with the quadrupole deformation parameter $\beta$ and neutron
occupation probability $P_{1s}(n)$. The bubble parameter decreases as a function of the
proton number and becomes negative (non-bubble) from $^{28}{\rm Si}$. 
We also see that, if we restrict the AMD calculation to spherical shape, $^{28}{\rm Si}$
also shows the bubble structure as displayed in Fig.~\ref{density.fig}~(e). This
confirms a strong impact of the nuclear  deformation on the bubble structure. 
With further increase of the proton number, in $^{30}$S, $^{32}$Ar and $^{34}$Ca, the
bubble structure in the matter density distributions are not seen since the central
densities of protons are already filled by the excess protons, while the neutron density
distributions still keep the bubble structure.

Thus, the present AMD calculation suggests
that $^{22}{\rm O}$ has both proton and
neutron bubble structure. We note, however,
this is in contradiction to the conclusion of
the mean-field calculations \cite{Grasso}.
It was shown that the bubble structure of
$^{22}{\rm O}$ is rather model dependent,
and pairing correlation tends to diminish the
bubble structure as it increases the neutron occupation of $1s$ orbit.
Since the present AMD calculation does not handle
the pairing correlation explicitly,
the stability of the bubble structure
of $^{22}{\rm O}$ shown in Fig.~\ref{density.fig}
needs to be investigated. To check the reliability of the AMD densities,
we calculated the mirror nucleus of $^{34}{\rm Ca}$, i.e., $^{34}{\rm Si}$
for which many calculations predicted proton
bubble structure~\cite{Grasso, Duguet17}
and indirect experimental evidence was obtained~\cite{mutschler}.
The calculated density shown in panel (i) of Fig.~\ref{density.fig} clearly
exhibits the proton bubble structure, which is very similar to that of
$^{34}{\rm Ca}$ and also that obtained
by the mean-field calculations~\cite{Grasso}.
Therefore, we conclude that the proton (neutron) bubble structure
of $^{34}{\rm Si}$ ($^{34}{\rm Ca}$ ) is robust,
while the bubble structure of $^{22}{\rm O}$ is somewhat model dependent.

\section{Discussions}
\label{discussion.sec}
\subsection{Extraction of bubble parameters for $N=14$ isotones}

In the previous section, we saw that $N=14$ isotones show remarkable variations in
their nuclear density profiles with  $^{22}$O exhibiting the most prominent bubble
structure, although the strong model dependence
was reported~\cite{Grasso}. We now examine
the possibility of extracting the degree of the bubble 
structure from the   reaction observables by performing a numerical test as
follows. First, we calculate the elastic scattering differential cross sections using
the density distributions obtained by the AMD, which we regard as the experimental data
(mock-up data). Then, by assuming spherical HO type density distributions defined in
Eq.~(\ref{HOdens.eq}), we fit the $\alpha$ (the mixing parameter) and size parameter of 
the HO to reproduce the position and magnitude of the first diffraction peak of the mock-up data. 
This procedure uniquely determines the spherical HO type density distribution from which we extract the bubble parameter $G$.
Thus, the obtained bubble parameter $G$ is compared with that of the
original one, $G_{\rm AMD}$ listed in Table~\ref{tab:deformation}, to test the
feasibility of the method.  

\begin{figure}[h!]
\centering
\includegraphics[width=\columnwidth]{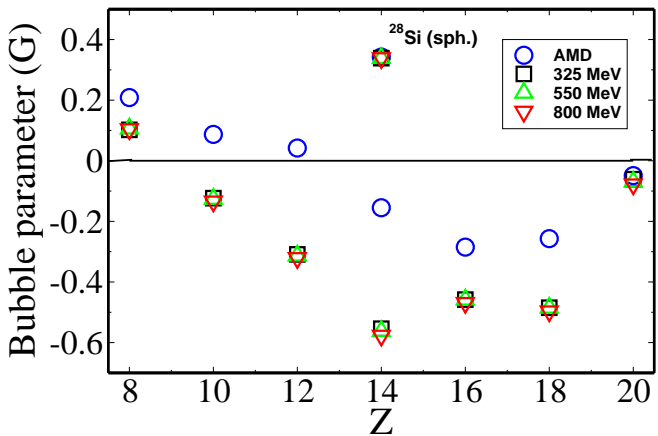}
\caption{Comparison of the bubble parameters obtained
  directly from the AMD densities and the elastic scattering analysis
  at the incident energies of 325, 550 and 800 MeV.}\label{bubble.fig}
\end{figure}

Figure~\ref{bubble.fig} plots the bubble parameters of $N=14$ isotones obtained 
from the mock-up data at the incident energies of 325 MeV, 550 MeV and 800 MeV, in
comparison with $G_{\rm AMD}$. It is noted that all the mock-up data (total reaction
cross sections calculated with AMD densities) are reproduced within 1\% differences.
The differences of the extracted bubble parameters are also less than 1\% for all the
incident energies. These show the robustness of this analysis.  
Although the bubble parameters extracted from the elastic scattering cross sections 
always undershoot the ``exact'' bubble parameters $G_{\rm AMD}$  (overestimate the
bubble structure), we do notice similarity in their behavior as a function of the
proton number. The disagreement is apparently due to the inappropriate assumption of the
model density - we assumed  spherical HO density distributions for all $N=14$
isotones. However, most of the nuclei are deformed inducing 
some deviations in the bubble parameter extraction. In fact, as we see in the
Fig.~\ref{bubble.fig}, the bubble parameter of $^{28}$Si is perfectly reproduced when we
constrain the AMD calculation to the spherical configuration. We also see a reasonable
description of almost spherical nuclei,  $^{22}$O ($Z=8$) and $^{34}$Ca ($Z=20$). Although it is beyond the scope of this paper, an analysis with more elaborated model density distributions including such as nuclear deformation is worth considering to obtain more precise determination of the bubble parameters. 
	
\subsection{How effective is proton scattering in probing the nuclear bubble?}

One may think that proton scattering
does not probe the nuclear bubble structure but only probes
the nuclear surface regions, the nuclear diffuseness.
To address this self-criticism,
we performed the same analysis as in the previous section
but with the two-parameter Fermi (2pF) model density,
$\rho_0/[1+\exp\left(r-R)/a\right]$, whose parameters $(R,a)$
are fixed so as to reproduce the first peak position
and its magnitude in the elastic scattering differential cross sections.
$\rho_0$, the central density, gets fixed from the normalization of the density distribution.
Obviously, the 2pF distribution has no bubble.
Note that with this analysis the 2pF model density
nicely reproduced the density profile at around the nuclear
surface of the realistic density distributions
obtained from the microscopic mean-field model~\cite{Hatakeyama18}.

\begin{figure}[ht]    
	\begin{center}    
		\includegraphics[width=\linewidth]{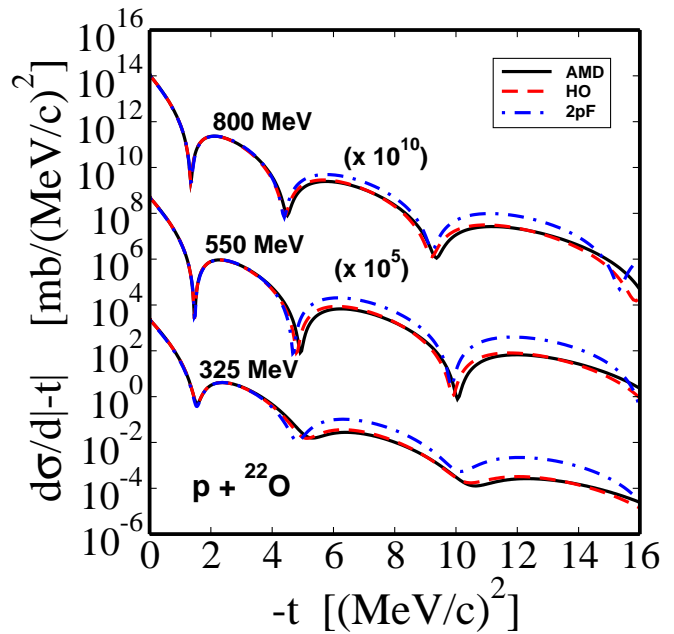}
		\caption{Elastic scattering differential cross sections
			of $p+^{22}$O with the AMD, HO, and 2pF densities.}
		\label{dcs22O.fig} 
	\end{center}
\end{figure}

Figure~\ref{dcs22O.fig} displays
the proton-$^{22}$O elastic scattering differential cross sections
with the AMD, HO and 2pF model densities
as a function of the four momentum transfer $|-t|$
at the incident energies of 325, 550, and 800 MeV.
The cross sections are essentially the same
up to $|-t|\approx 3$ GeV/$c$, which is understandable as both the HO
and 2pF model densities are adjusted to reproduce the position and magnitude
of the first diffraction peak. However, beyond this limit while the HO
and AMD results agree with each other, those with the 2pF model
density deviate significantly.

We already know in Fig.~\ref{reacprob.fig}
the fact that the incident proton cannot probe differences in
the internal densities below $\approx 2$ fm.
The difference of the density profiles
in the middle to the surface regions, in which
the bubble structure is still not masked, can be distinguished
by analyzing the cross sections in the backward angles
beyond the first peak. The proton-nucleus scattering can indeed be 
an effective tool to probe the bubble structure in exotic nuclei.
We remark that similar indication was found in the analysis of
proton-$^{48}$S scattering with bubble and non-bubble density
profiles~\cite{Furumoto19}.

\section{Conclusions}
\label{conclusions.sec}
Nuclei with a depression in the central part of their density
- the so-called bubble structure - has attracted attention in recent times. 
Considerable efforts are underway to look for suitable probes
for these exotic systems.
In this work, we have discussed the feasibility of using
a proton probe to extract the degree of the bubble structure.
We have calculated the structure of even-even $N=14$ isotones in the 
$22\leq A \leq 34$ mass range
using a microscopic structure model,
the antisymmetrized molecular dynamics (AMD).
The Glauber model is then employed to evaluate
reaction observables of high-energy nucleon-nucleus scattering.

Due to the strong absorption in the internal region of the target nucleus,
the bubble structure or the central depression of the target density,
cannot be directly measured using the proton probe. However, effects
of this structure are reflected from the middle to the surface regions
of the nuclear density. They also tend to have a sharper nuclear surface.
Furthermore, nuclear deformation acts as an hindrance to the emergence
of the bubble structure.

We find that the AMD calculation predicts prominent bubble structure of
 $^{22}$O, which exhibits a small deformation, after analyzing a
host of $N=14$ isotones.
The degree of the bubble structure is extracted
by a systematic analysis of the calculated cross sections
obtained with the AMD by using
simple harmonic-oscillator type model densities.
To improve the accuracy of the extraction,
it is necessary to employ a more realistic model density
that can describe the nuclear deformation.

We have shown that the bubble structure information
is imprinted on the nucleon-nucleus elastic scattering
differential cross sections and is possibly extracted by
analyzing the cross sections up to the first diffraction peak.
Nevertheless a more accurate analysis involving the second
diffraction peak would be a welcome addition. 

\acknowledgments
This work was supported by  
JSPS KAKENHI Grants Nos. 18K03635, 18H04569, 19H05140,
and 19K03859, the Collaborative Research Program 2020,
Information Initiative Center, Hokkaido University and the
Scheme for Promotion of Academic and Research Collaboration
(SPARC/2018-2019/P309/SL), MHRD, India. V.C. also acknowledges MHRD,
India for a doctoral fellowship and a grant from SPARC to visit
the Hokkaido University.

\appendix

 \section{Estimation of the deformation parameters and occupation probabilities from the
 AMD wave functions} \label{app:amd}
 Here, we explain how we estimated the quadrupole deformation parameters
 and single-particle occupation probabilities
 of the AMD wave functions for $N=14$ isotones listed in 
 Table~\ref{tab:deformation}.
 The AMD wave function given in Eq.~(\ref{eq:gcmwf}) is, 
 in general, a superposition of the Slater determinants
 with different deformation and
 different single-particle configurations.
 Therefore, to estimate these quantities, we
 pick up the Slater determinant $\Phi(\beta)$, which has the maximum
 overlap with the AMD wave function  $|\braket{P^{J=0}\Phi(\beta)|\Psi_0}|$,
 and regard it as an approximation of the AMD wave function $\Psi_0$. 

 The deformation parameters $\beta$ and $\gamma$ of $\Psi_0$
 may be approximated by those of  $\Phi(\beta)$.
 The occupation probabilities of the $1s$-orbit are also estimated in
 a similar manner. We calculate the single-particle energies
 and orbits of described by $\Phi(\beta)$
 by using the AMD+HF method~\cite{dote}. Because of the nuclear deformation,
 the single-particle orbits, $\phi_1(\bm r),...,\phi_A(\bm r)$,
 are no longer the eigenstates of the angular momentum. Therefore,
 we consider the multipole decomposition of them, 
\begin{align}
  \phi_i(\bm r) =
  \sum_{jlj_z}\phi_{i;jlj_z}(r)\left[Y_l(\hat r)\times \chi_{1/2}\right]_{jj_z}.
\end{align}
The squared amplitudes for the $j=1/2$ and $l=0$ components should
give us an estimate of the occupation probability.
Assuming the complete filling of the $0s$-orbit, the
neutron ($n$) and proton ($p$)
occupation probabilities of the $1s$-orbit are obtained
approximately as
\begin{align}
  P_{1s}(n/p) = \sum_{i=1}^{N/Z}\sum_{j_z=-\frac{1}{2}}^{\frac{1}{2}}
  \left|\left<\phi_{i;\frac{1}{2}0j_z}\right|
  \left.\phi_{i;\frac{1}{2}0j_z}\right>\right|^2 - 2.
\end{align}



\begin{thebibliography}{99}
\bibitem{Tanihata85}
  I. Tanihata, H. Hamagaki, O. Hashimoto, Y. Shida, N. Yoshikawa, K. Sugimoto,
  O. Yamakawa, T. Kobayashi, and N. Takahashi,
  Phys. Rev. Lett. {\bf 55}, 2676 (1985).
\bibitem{Tanihata13}
  I. Tanihata, H. Savajols, and R. Kanungo,
  Prog. Part. Nucl. Phys. {\bf 68}, 215 (2013).
\bibitem{Hofstadter}
R. Hofstadter, Rev. Mod. Phys. \textbf{28}, 214 (1956).
\bibitem{Grasso}
M. Grasso, L. Gaudefroy, E. Khan, T. Nik\v{s}i\'{c},
J. Piekarewicz, O. Sorlin, N. Van Giai, and D. Vretenar,
  Phys. Rev. C \textbf{79}, 034318 (2009).
\bibitem{Li}
  J.~J. Li, W.~H. Long, J.~L. Song, and Q. Zhao,
  Phys. Rev. C \textbf{93}, 054312 (2016).	
\bibitem{Campi}
  X. Campi and  D.~W.~L. Sprung,
  Phys. Lett. \textbf{B 46}, 291 (1973).
\bibitem{Davis}
  K.~T.~R. Davis, S.~J. Krieger,
  and C. Y. Wong, Nucl. Phys. \textbf{A 216}, 250 (1973).
\bibitem{Khan}
  E. Khan, M.  Grasso, J.  Margueron, and N. Van Giai,
  Nucl. Phys. \textbf{A 800}, 37  (2008).
\bibitem{Bender}
  M. Bender, K. Rutz, P.-G. Reinhard, J.~A. Maruhn, and W. Greiner,
  Phys. Rev. C \textbf{60}, 034304 (1999).  
\bibitem{mutschler}
  A. Mutschler, A. Lemasson, O. Sorlin, D. Bazin, C. Borcea,
  R. Borcea, Z. Dombr\'adi, J.-P. Ebran, A. Gade, H. Iwasaki {\it et al. },
Nature Physics \textbf{13}, 152 (2017).
\bibitem{Tsukada} K. Tsukada, A. Enokizono, T. Ohnishi, K. Adachi, T. Fujita,
   M. Hara, M. Hori, T. Hori, S. Ichikawa, K. Kurita {\it et al}.,
         Phys. Rev. Lett. {\bf 118}, 262501 (2017).
\bibitem{PREX}
  S. Abrahamyan {\it et al.} (PREX collaboration),
  Phys. Rev. Lett. {\bf 108}, 112502 (2012).	
\bibitem{Sakaguchi}
  H. Sakaguchi and J. Zenihiro, Prog. Part. Nucl. Phys.
  \textbf{97}, 1 (2017), and references therein
\bibitem{Matsuda13} Y. Matsuda, H. Sakaguchi, H. Takeda, S. Terashima,
  J. Zenihiro, T. Kobayashi, T. Murakami, Y. Iwao, T. Ichihara,
  T. Suda {\it et al.},
 Phys. Rev. C {\bf 87}, 034614 (2013).
\bibitem{glauber} R.~J. Glauber, {\it Lectures in Theoretical Physics},
  edited by W.~E. Brittin and L.~G. Dunham
  (Interscience, New York, 1959), Vol. 1, p.315.             
\bibitem{Varga02} K. Varga, S.~C. Pieper, Y. Suzuki, and R.~B. Wiringa,
  Phys. Rev. C {\bf 66}, 034611 (2002).
\bibitem{Nagahisa18} T. Nagahisa and W. Horiuchi,
  Phys. Rev. C {\bf 97}, 054614 (2018).		
\bibitem{Ibrahim09} B. Abu-Ibrahim, S. Iwasaki, W. Horiuchi, A. Kohama,
  and Y. Suzuki, J. Phys. Soc. Jpn., Vol. {\bf 78}, 044201 (2009).  
\bibitem{Hatakeyama18} S. Hatakeyama, W. Horiuchi, and A. Kohama,
Phys. Rev. C {\bf 97}, 054607 (2018).
\bibitem{Lray}L. Ray, Phys. Rev. C \textbf{20}, 1857 (1979).
\bibitem{Horiuchi07} W. Horiuchi, Y. Suzuki, B. Abu-Ibrahim,
  and A. Kohama, Phys. Rev. C {\bf 75}, 044607 (2007).
\bibitem{Ibrahim08} B. Abu-Ibrahim, W. Horiuchi, A. Kohama, and Y. Suzuki, 
  Phys. Rev. C {\bf 77}, 034607 (2008);
  {\it ibid} {\bf 80}, 029903(E) (2009);
  {\bf 81}, 019901(E) (2010).                
\bibitem{Angeli13} I. Angeli, K.~P. Marinova,
  At. Data Nucl. Tables {\bf 99}, 69 (2013).		
\bibitem{Amado}
  R.~D. Amado, J. P. Dedonder, and F. Lenz,
  Phys. Rev. C \textbf{21}, 647 (1980).              
\bibitem{Kohama04} A. Kohama, K. Iida, and K. Oyamatsu,
  Phys. Rev. C {\bf 69}, 064316 (2004).
\bibitem{Kohama05} A. Kohama, K. Iida, and K. Oyamatsu,
  Phys. Rev. C {\bf 72}, 024602 (2005).
\bibitem{Kohama16} A. Kohama, K. Iida, and K. Oyamatsu,
  J. Phys. Soc. Jpn. {\bf 85}, 094201 (2016).
\bibitem{AMDRev1} Y. Kanada-Enyo, M. Kimura, and H. Horiuchi,
  Comptes Rendus Physique {\bf 4}, 497 (2003).
\bibitem{AMDRev2} M. Kimura, Phys. Rev. C {\bf 69}, 044319 (2004).
\bibitem{GognyD1S}  J. Berger, M. Girod, and D. Gogny,
  Comp. Phys. Comm. {\bf 63}, 365 (1991). 
\bibitem{Sumi12} T. Sumi, K. Minomo, S. Tagami, M. Kimura, T. Matsumoto,
  K. Ogata, Y.~R. Shimizu, and M. Yahiro,
  Phys. Rev. C {\bf 85}, 064613 (2012).
\bibitem{Wata14} S. Watanabe, K. Minomo, M. Shimada, S. Tagami, M. Kimura,
  M. Takechi, M. Fukuda, D. Nishimura, T. Suzuki, T. Matsumoto,
  Y.~R. Shimizu, and M. Yahiro, Phys. Rev. C {\bf 89}, 044610 (2014). 
\bibitem{Peru14} S. P\`{e}ru and  M. Martini,
  Eur. Phys. J. A{\bf 50}, 88 (2014).
\bibitem{cooling} Y. Kanada-Enyo, M. Kimura, and A. Ono,
  Prog. Theor. Exp. Phys. {\bf 2012}, 1A202 (2012).
\bibitem{Hill53} D. L. Hill and J. A. Wheeler,
	Phys. Rev. {\bf 89}, 1102 (1953).
 \bibitem{Duguet17} T. Duguet, V. Som\`a, S. Lecluse, C. Barbieri, and P. Navr\'atil
	 Phys. Rev. C{\bf 95}, 034319 (2017).
\bibitem{Furumoto19} T. Furumoto, K. Tsubakihara, S. Ebata,
  and W. Horiuchi, Phys. Rev. C {\bf 99}, 034605 (2019).
\bibitem{dote} A. Dote, H. Horiuchi and  Y. Kanada-En'yo,
  Phys. Rev. C {\bf 56}, 1844 (1997).
\end{thebibliography}
\end{document}